# Effect of two bands on critical fields in MgB$_2$ thin films with various resistivity values

V.Ferrando*, P.Manfrinetti$^§$, D.Marré*, M.Putti*, I.Sheikin$^{§§}$, C.Tarantini*, and C.Ferdeghini*

* *INFM-LAMIA, Dipartimento di Fisica, Università di Genova, via Dodecaneso 33, 16146 Genova Italy*

$^§$ *INFM-LAMIA, Dipartimento di Chimica e Chimica Industriale, via Dodecaneso 31, 16146 Genova, Italy*

$^{§§}$ *GHMFL, MPI-FKF/CNRS, 28 Avenue des Martyrs, BP 166, 38042 Greboble Cedex 9, France*

**Abstract**

Upper critical fields of four MgB$_2$ thin films were measured up to 28 Tesla at Grenoble High Magnetic Field Laboratory. The films were grown by Pulsed Laser Deposition and showed critical temperatures ranging between 29.5 and 38.8 K and resistivities at 40 K varying from 5 to 50 μΩcm. The critical fields in the perpendicular direction turned out to be in the 13-24 T range while they were estimated to be in 42-57 T the range in *ab*-planes. In contrast to the prediction of the BCS theory, we did not observe any saturation at low temperatures: a linear temperature dependence is exhibited even at lowest temperatures at which we made the measurements. Moreover, the critical field values seemed not to depend on the normal state resistivity value. In this paper, we analyze these data considering the multiband nature of superconductivity in MgB$_2$ We will show how the scattering mechanisms that determine critical fields and resistivity can be different.

# 1. INTRODUCTION

Since the discovery of superconductivity in magnesium diboride[1], several unusual properties arising from the presence of two distinct s-wave gaps have been emphasized. It has been clarified[2,3] that two different gaps are associated with two distinct sheets of Fermi surface. The larger gap is associated with the σ bands, while the smaller to the π bands. π and σ bands have different characteristics, π-bands being essentially electron type and nearly isotropic and σ-bands essentially hole-type and nearly two dimensional; σ-bands determine the anisotropy of physical properties.

Due to the different parity of the σ and π bands, the inter-band impurity scattering is expected to be negligible compared with the intra-band ones; thus, σ and π bands can be considered as different channels conducting in parallel. This scenario gives an explanation of some superconducting properties[4-6], but the effect of the presence of the two bands on the critical fields is still not clear.

Upper critical fields and their anisotropy can be studied on single crystals or on *c*-axis oriented (or better epitaxial) thin films. Important differences exist between these two kinds of samples. Single crystals present homogeneous optimal $T_c$ values, low residual resistivity $\rho_0$ (about 2-5 μΩcm), relatively low critical fields values perpendicular ($H_{c2}^\perp$) and parallel ($H_{c2}^\parallel$) to the ab-planes ($H_{c2}^\perp$ = 3-5 T and $H_{c2}^\parallel$ = 16-19 T) and an anisotropy factor $\gamma = H_{c2}^\parallel / H_{c2}^\perp$ =5-6 always decreasing with increasing temperature[7-10]. On the contrary, thin films show an important spread in $T_c$ and $\rho_0$ values. $T_c$ can vary from the optimal value down to 25 K and resistivity from few μΩcm up to hundred of μΩcm. In thin films the critical field values are considerably higher (up to tens of Teslas) and γ values are always lower (up to 3.5); γ usually decreases when temperature increases, even though in some cases the opposite behavior was also observed[11-20].

The difference between the properties of single crystals and thin films can be ascribed to disorder, which is surely stronger in thin films. Disorder can play a role in suppressing $T_c$ and in increasing the critical fields: in a BCS scenario, the critical field can be enhanced by increasing the resistivity, but it is still not clear how thin films with low resistivity can show very high critical field, as in the case reported here.

This paper presents data relevant to four thin films with resistivity values ranging within one order of magnitude. Our goal is to focus the role of disorder in thin films in order that the relations among $\rho_0$, $T_c$, and critical fields may be clarified. First, we study the resistivity curves in detail to estimate the more important scattering mechanisms in our films. Second, the critical field data are analyzed within the Gurevich model[21], which correlates the critical fields to the diffusivity of each band. Finally, the scattering mechanisms determining resistivity and critical field values are compared.

## 2. SAMPLE PREPARATION AND CHARACTERIZATION

In order to study the influence of disorder on the upper critical field behavior of $MgB_2$, we have measured four different films prepared by standard two-step method[11] on different substrates. The samples, whose thickness is in the range 900-1300 Å, were deposited by pulsed laser ablation starting from stoichiometric target; details about the deposition technique are reported elsewhere[22]. In the following, they will be referred to as film 1, film 2, film 3 and film 4; their properties are summarized in Table 1. The properties of these films vary from film 1, which presents low critical temperature (29.5 K) and low residual resistivity ratio (RRR = 1.2), to film 4, which shows $T_c$ = 38.8 K, near to the bulk value, and a relatively high RRR (2.5). In figure 1 the resistivity versus temperature curves are plotted. Normal state resistivity is related to the different purity of the samples; in fact, just above the transition, resistivity values ranging between 50 $\mu\Omega$cm and 5 $\mu\Omega$cm have been found. It should be noted that in these samples the resistivity at 40 K increases of one order of magnitude while the change in resistivity $\Delta\rho=\rho(300K)-\rho(40K)$ remains nearly constant. All the samples show good structural properties, as evidenced by X-ray diffraction measurements. In all $\theta$-$2\theta$ patterns, intense (00l) peaks coming from $MgB_2$ can be detected, indicating a strong *c*-axis orientation of the phase. Only in film 4, (101) reflection, which is the most intense in powders, seems to be detectable, even though with very low intensity; this indicates a not perfect orientation of the film. We have already reported[15] that samples with critical temperature near to the optimal value often present the worst structural properties, while samples with low $T_c$ and RRR values are usually more oriented and sometimes show in plane texturing and very high critical fields. From $\phi$ scan measurements, we had clear indications of in plane alignment for film 1[19]. Up to now, similar measurements have not been performed on the other films. From a structural point of view, a dependence of the cell parameters on the substrate used has also been observed. In particular, the *c* parameter, calculated from the position of (002) peak, seems to be smaller than the optimal value in films grown on *c*-cut sapphire (in our case film 1 and film 4), while it is slightly higher in samples deposited on (111) MgO (film 2 and film 3). This was verified in all the films grown on these two kind of substrates.

## 3. NORMAL STATE RESISTIVITY

To analyze the scattering mechanisms in our films we consider the normal state electrical resistivity. We recall that thin films, generally, show higher resistivity and lower residual resistivity ratio compared with single crystals, because of the high structural disorder and nanostructure, which can induce grain boundaries scattering. Nowadays, thin films with resistivity curves very similar to those of single crystals and residual resistivity $\rho_0$ of the order of few $\mu\Omega$cm are available, and film 4 is one of them. In particular, resistivity of film 4 follows the power law $\rho(T)=\rho_0+\alpha T^3$ up to 100 K, as usually occurs in $MgB_2$ single crystals and bulk samples[24].

In Table I some data drawn from the resistivity curves have been summarized: the resistivity at 40 K (in the following considered to be nearly equal to the residual resistivity $\rho_0$), the resistivity slope calculated at room temperature, $d\rho/dT(300\ K)$, and the residual resistivity ratio. We point out that the first two values, owing to the uncertainty in the film thickness evaluation, have an uncertainty of 20%, but the following discussion is not affected by such indetermination.

If the inter-band scattering rate is negligible compared with the intra-band ones, $\rho_0$ is given by the parallel of $\rho_{0\sigma}$ and $\rho_{0\pi}$, the residual resistivity of $\sigma$ and $\pi$ bands, respectively, i.e.:

$$\frac{1}{\rho_0}=\frac{1}{\rho_{0\sigma}}+\frac{1}{\rho_{0\pi}}=\omega_{p\sigma}^2\varepsilon_0\frac{1}{\Gamma_{\sigma\sigma}}+\omega_{p\pi}^2\varepsilon_0\frac{1}{\Gamma_{\pi\pi}} \qquad (1)$$

Here, $\omega_{p\sigma}$ and $\omega_{p\pi}$ are the plasmon frequencies of $\sigma$ and $\pi$ bands, $\varepsilon_0$ the vacuum dielectric constant and $\Gamma_{\sigma\sigma}$ and $\Gamma_{\pi\pi}$ are the intra-band scattering rates. Since in the ab-plane $\omega_{p\sigma}$ and $\omega_{p\pi}$ do not differ too much (4.14 and 5.89 eV, respectively[5]), we can define an average plasmon frequency $\omega_p^2=(\omega_{p\sigma}^2+\omega_{p\pi}^2)/2$; thus, from the residual resistivity we can calculate the parallel between $\Gamma_{\sigma\sigma}$ and $\Gamma_{\pi\pi}$:

$$\Gamma=\frac{\Gamma_{\sigma\sigma}\Gamma_{\pi\pi}}{\Gamma_{\sigma\sigma}+\Gamma_{\pi\pi}}=\varepsilon_0\rho_0\omega_p^2 \qquad (2)$$

The calculated $\Gamma$ values are reported in Table II and vary from 171 to 17meV.

These values represent the effective scattering in each film but we have to clarify which band is more affected. To this purpose we analyse the resistivity slope. In fact, the resistivity slope can change depending on whether $\sigma$ or $\pi$ conduction band prevails[23, 24]. The resistivity slope for $\sigma$ and $\pi$ bands, $d\rho_\sigma/dT$ and $d\rho_\pi/dT$, are given by [23]:

$$\frac{d\rho_\sigma}{dT} = \frac{1}{\varpi_{p\sigma}^2 \varepsilon_0} \frac{2\pi k_B \lambda_{tr\sigma}}{\hbar} \approx 0.26 \,\mu\Omega\text{cm/K}$$

$$\frac{d\rho_\pi}{dT} = \frac{1}{\varpi_{p\pi}^2 \varepsilon_0} \frac{2\pi k_B \lambda_{tr\pi}}{\hbar} \approx 0.06 \,\mu\Omega\text{cm/K}$$

where $\lambda_{tr\sigma}$ and $\lambda_{tr\pi}$ (1.1 and 0.56, respectively[25]) are the transport electron-phonon coupling constants.

Due to the lower coupling constant and to the larger plasmon frequency, the phonon contribution to resistivity is lower for $\pi$ band. In clean samples this contribution prevails and at room temperature a slope close to $d\rho_\pi/dT$ is expected. On the other hand, in dirty samples the value of the resistivity slope depends on the ratio between the residual resistivities of $\sigma$ and $\pi$ bands, $\rho_{0\sigma}/\rho_{0\pi}$; when $\rho_{0\sigma}/\rho_{0\pi} \gg 1$, $d\rho/dT \approx d\rho_\pi/dT$ while when $\rho_{0\sigma}/\rho_{0\pi} \ll 1$, $d\rho/dT \approx d\rho_\sigma/dT$. Looking up the $d\rho/dT$(300 K) values[26] of Table I, one can see that the slopes of the films are close to the $d\rho_\pi/dT$ value; only film 2 has an intermediate slope between $d\rho_\sigma/dT$ and $d\rho_\pi/dT$, but however closer to $d\rho_\pi/dT$.

In conclusion, in the films here presented, the $\pi$ conduction prevails and so we can assume $\rho \approx \rho_\pi < \rho_\sigma$ and $\Gamma \approx \Gamma_{\pi\pi} < \Gamma_{\sigma\sigma}$; this could be due to disorder, especially effective in the B-planes. This result has to be considered in the following analysis of critical fields data.

Really, the analysis of resistivity data as a tool to extract information on multiband effects in $MgB_2$ has been questioned by Rowell[27]. In his paper, he showed how grain boundary scattering and poor connectivity between grains can make the actual geometrical factor for the calculation of resistivity hard to estimate. Due to this uncertainty, the calculated resistivity, as well as its variation, can be overestimated. Even if this overestimation is present in our data, the actual $d\rho/dT$ values would become even lower, therefore reinforcing our previous conclusions.

Finally, we suggest that disorder could also be the cause of the $T_c$ suppression in thin films. In our case in fact, this suppression seems not to be caused by uniaxial stresses. As already mentioned in the previous paragraph, the *c*-axis values depend on the kind of substrate but no correlation between the lattice parameters and the critical temperature has been observed. In a two-gap superconductor in the absence of magnetic scattering, only the inter-band scattering rate, $\Gamma_{\sigma\pi}$, is able to reduce the critical temperature[24, 25]; the equation which describes the $T_c$ suppression in the case of $MgB_2$ is given by[24]:

$$\left(\frac{\delta T_c}{T_c}\right) \approx -\frac{\pi \Gamma_{\sigma\pi}}{8 k_B T_c} \frac{(\Delta_\sigma - \Delta_\pi)(\Delta_\sigma N_\pi - \Delta_\pi N_\sigma)}{(\Delta_\sigma^2 + \Delta_\pi^2) N_\pi} \qquad (3)$$

where $\delta T_c$ is the critical temperature reduction with respect to the optimal value, $\Delta_\sigma$ and $\Delta_\pi$ are the gap amplitudes at T = 0 K and $N_\sigma$ and $N_\pi$ are the density of states of σ and π bands, respectively.

By introducing the following values: $\Delta_\sigma$ = 7 meV, $\Delta_\pi$ =2.2 meV, $N_\sigma$ = 0.3 states/eVcell and $N_\pi$ = 0.4 states/eV cell [2] in eq. (1), we find:

$$\left(\frac{\delta T_c}{T_c}\right) \approx -0.2 \frac{\Gamma_{\sigma\pi}}{k_B T_c} \qquad (4)$$

If we assume an optimal $T_c$ value of 39 K, we can calculate $\Gamma_{\sigma\pi}$ for each film: the values range from 0.1 to 4 meV and are reported in Table II. We recall that, owing to the different parity of σ and π bands, $\Gamma_{\sigma\pi}$ is expected to be very low and, in general, negligible compared with $\Gamma_{\sigma\sigma}$ and $\Gamma_{\pi\pi}$. $\Gamma_{\sigma\pi}$ values, compared with the intra-band scattering rates estimated before, turn out to be more than one order of magnitude lower for all the films, even for film 1, which presents a conspicuous $T_c$ suppression (10 K). Moreover, a rough correlation between intra and inter-band scattering rates can be observed: the latter increase as far as the first increase.

We conclude that the large spreading of $T_c$ values observed in thin films rather than in bulk samples can be caused by the large structural disorder presented by films. In any case the condition $\Gamma_{\sigma\pi} \ll \Gamma_{\sigma\sigma}$, $\Gamma_{\pi\pi}$ is fairly met.

## 4. UPPER CRITICAL FIELD

High magnetic field electrical resistance measurements up to 28 T and down to 2 K were performed at GHMFL (Grenoble High Magnetic Field Laboratory) using a standard four-probe AC resistance technique. For each temperature upper critical field $H_{c2}^{\parallel}$ and $H_{c2}^{\perp}$ have been estimated as the point of the transition in which the resistance is 90% of the normal state value.

In figure 2, $H_{c2}^{\parallel}$ and $H_{c2}^{\perp}$ are reported as a function of the reduced temperature for the four samples. Despite the great difference in critical temperature and resistivity values, a common trend of $H_{c2}^{\perp}$(T) curves is exhibited by all the films but film 4 (the film with the lowest resistivity value), whose $H_{c2}^{\perp}$ seems to be considerably higher than the others (24 T instead of 14-16 T for the other three samples at 2 K). We recall that, as previously observed, this sample is not completely c-oriented: if upper critical fields are determined with the criterion of 90% of the normal state resistivity, the not aligned regions can only cause an overestimation of the smaller critical field ($H_{c2}^{\perp}$), the larger ($H_{c2}^{\parallel}$) being not affected. In fact misaligned grains remain superconductor at fields higher than $H_{c2}^{\perp}$ but, in any case lower than $H_{c2}^{\parallel}$. Therefore, a comparison of the four $H_{c2}^{\parallel}$ curves of figure 2 is possible. $H_{c2}^{\parallel}$ values at low temperature are quite similar and the derivative is even higher for film 4. Two interesting features can be noted in these data: first, the upward curvature near $T_c$, becoming more evident when the critical temperature value is near to the optimal one[12,19], and second the linearity of the $H_{c2}(T)$ curves, even at the lowest temperatures we measured (2 K in the case of perpendicular orientation).

In low $T_c$ superconductors in the dirty limit, the zero-temperature upper critical field can be calculated, in a simple BCS framework, from

$$H_{c2}(0) = 0.69 T_c \left( \frac{dH_{c2}}{dT} \right) \qquad (5)$$

with

$$\frac{dH_{c2}}{dT} = \frac{4 e k_B}{\pi \hbar} N_F \rho_0 \qquad (6)$$

where $N_F$ is the density of states at the Fermi surface and $\rho_0$ the normal state residual resistivity. It should be noted that, in this case, this standard BCS formula fails. In fact this BCS extrapolation strongly underestimates the real $H_{c2}(0)$; for film 1, for example, we calculate the BCS zero-temperature extrapolation value of $H_{c2}^{\parallel}(0)=22T$ and $H_{c2}^{\perp}(0)=8.75$ Tesla, whereas these values have already been reached at 13 and 10 K, respectively. In contrast with the BCS theory predictions, we have not observed any saturation of $H_{c2}$ at low temperature: a linear temperature dependence is observed even at the lowest temperatures we measured. This is fairly evident, in particular, in perpendicular orientation, where the magnetic field we can apply is strong enough to determine $H_{c2}$ down to 2 K, which allows a reasonable estimation of $H_{c2}^{\perp}(0)$ by linear extrapolation. The obtained values are reported in Table II.

The slope $dH_{c2}/dT$ at $T_c$ is proportional to residual resistivity (see eq.(6)), so an increase in $\rho$ should proportionally increase the upper critical field values. This was verified for low temperature superconductors and represents the usual method to enhance $H_{c2}$ in technological materials, such as Nb-Ti and A15 compounds. The same approach was followed also for $MgB_2$, where the resistivity was increased both by alloying[12] and irradiating[28, 29] the phase, leading to a rise of critical fields in both cases. In the whole set of our data, on the contrary, we cannot observe a clear $H_{c2}$ dependence on ρ. In the four samples, the resistivity values just above the transition vary by one order of magnitude (from 5 to 50 μΩcm) but the critical fields values are similar to each other.

From the data of figure 2, it is possible to estimate the anisotropy factors $\gamma = H_{c2}^{\parallel}/H_{c2}^{\perp}$ for all the films. Their temperature dependences are shown in figure 3. All the γ(T) curves have the same behavior, γ decreasing with increasing temperature. At the lowest temperatures the anisotropies of all the films are in the range between 3 and 3.5 (the maximum value reported for γ up to now for films), the only exception being film 4 for which γ is 2.3, probably because of the not perfect c-orientation. The maximum γ value reported in the literature for films must be compared with 5-6 reported for single crystals. An understanding of this topic is still lacking in the literature. For our purpose, considering that the γ curves seem to saturate at low temperature, it is reasonable to use the γ values at the lowest temperature to estimate $H_{c2}^{\parallel}(0)$ from the $H_{c2}^{\perp}(0)$ values (see Table II). The so calculated parallel critical field values are also reported in Table II and range between 42 and 57 Tesla, which are values of great interest for high field application of superconductivity.

It is clear that, in modelling critical field behaviour, the two-band nature of superconductivity in $MgB_2$ has to be taken into account. Articles describing the critical field behaviour in this framework only recently began to appear in the literature[21, 30].

The model proposed by Gurevich[21] considers the intra-band electronic diffusivities $D_\pi$ and $D_\sigma$, the inter-band one being neglected; the upper critical field is determined by the smaller (or larger) one depending on the temperature range considered. The shape can be considerably different from the BCS one and $H_{c2}(0)$ can drastically exceed the BCS extrapolation.

For $H_{c2}(0)$ the following equation has been given[21]:

$$H_{c2}(\eta) = \frac{k_B \phi_0 T_c e^{\frac{g(\eta)}{2}}}{2\hbar \nu D_\sigma \sqrt{\eta}} \qquad (7)$$

with

$$g(\eta) = \frac{1}{2}\left( \sqrt{\ln^2(\eta) + 2\frac{\lambda_m \ln(\eta)}{w} + \frac{\lambda_0^2}{w^2}} - \frac{\lambda_0}{w} \right) \qquad (8)$$

where $\lambda_m = \lambda_{\sigma\sigma} - \lambda_{\pi\pi}$, $\lambda_0 = (\lambda_m^2 + 4\lambda_{\sigma\pi}\lambda_{\pi\sigma})^{1/2}$, $\ln \nu = -0.577$, $w = \lambda_{\sigma\sigma}\lambda_{\pi\pi} - \lambda_{\sigma\pi}\lambda_{\pi\sigma}$ and $\eta = D_\pi/D_\sigma$.

Eq.(7) can be specialized for the three different conditions $\eta\gg1$, $\eta\ll1$ and $\eta=1$ giving:

$$H_{c2}(0) = \frac{\phi_0 k_B T_c}{2\hbar v\, D_\sigma} e^{-\lambda_2/2w} \qquad \text{for } D_\sigma \ll D_\pi \ (\eta\gg1) \qquad (9a)$$

$$H_{c2}(0) = \frac{\phi_0 k_B T_c}{2\hbar v\, \sqrt{D_\sigma D_\pi}} \qquad \text{for } D_\sigma = D_\pi \ (\eta=1) \qquad (9b)$$

$$H_{c2}(0) = \frac{\phi_0 k_B T_c}{2\hbar v D_\pi} e^{-\lambda_1/2w} \qquad \text{for } D_\pi \ll D_\sigma \ (\eta\ll1) \qquad (9c)$$

with $\lambda_{1,2} = \lambda_0 \pm \lambda_m$. Interestingly, the zero-temperature upper critical field value is always dominated by the lowest diffusivity when $D_\sigma$ and $D_\pi$ are different, and by the geometrical media when they are similar. The intermediate case is similar to the BCS one. What marks the three different conditions is the dependence of critical fields anisotropy on temperature: if $D_\sigma \ll D_\pi$, $\gamma$ increases when temperature decreases, while the temperature dependence is the opposite if $D_\pi \ll D_\sigma$. For $D_\sigma \sim D_\pi$, finally, $\gamma$ is nearly constant and only a slight increase is observed as temperature decreases.

In the framework of the Gurevich model, the $\gamma$ temperature dependences of figure 3 seem to indicate that we are in the $D_\pi \geq D_\sigma$ condition. This is in agreement with our results on normal state resistivity: in fact, we found $\rho_\pi < \rho_\sigma$ for all the films, which implies $D_\pi > D_\sigma$ ($\eta>1$). With this hypothesis on the diffusivity ratio, it is possible to estimate $D_\sigma$ from the measured $H_{c2}^\perp(0)$. In fact, for $\eta>1$ eq.(7) depends weakly on $\eta$ and the calculated $D_\sigma$ values vary only by 4% as $\eta$ varies from 1 to 10. The obtained $D_\sigma$ values are reported in Table II and they are similar for film 1, 2 and 3 (around $0.48 \cdot 10^{-3}$ m$^2$s$^{-1}$) and slightly lower only for film 4, which presents higher $H_{c2}^\perp$ value. Using these $D_\sigma$ values, the resistivity associated to the $\sigma$-bands $\rho_\sigma$ can be calculated from $\frac{1}{\rho_\sigma} = e^2 N_\sigma D_\sigma$ (see Table II).

$\rho_\sigma$ ranges between 123 and 163 $\mu\Omega$cm, values considerably higher compared with the measured $\rho_0$ (5-50 $\mu\Omega$cm).

This is an important result: in a two band superconductor the resistivity and the critical field can be determined by two different mechanisms (in our case the scattering in the $\sigma$-bands for critical field and the scattering in the $\pi$-bands for resistivity). From the comparison between the measured $\rho_0$ and $\rho_\sigma$, it is possible to estimate $\rho_\pi$: it varies between 5 and 83 $\mu\Omega$cm, as reported in Table II. From these values we calculate $D_\pi$ and, finally, we could estimate $\eta$. It turned to be 1.6,

1.1, 4.2 and 24 for film 1, 2, 3 and 4, respectively, which confirms that the assumptions made were reasonable. Finally we have found that the critical fields of our films are determined by the diffusivity of σ band, which is the lower, while resistivities are determined by the larger diffusivity, $D_\pi$. This explain why it is possible to have high critical fields in low resistivity films. What is peculiar in our films is that their $D_\sigma$ is nearly the same (in fact, they have similar critical fields) but $D_\pi$ changes by more than one order of magnitude. This could be due to the fact that disorder in the B-planes that forms in the course of the deposition process is poorly recovered during annealing in Mg atmosphere for the phase crystallization.

## 5.CONCLUSIONS

We studied the role of disorder in thin films with different values of resistivity and critical temperature, but similar values of critical fields. We suggest that the Tc suppression is determined by the inter-band impurity scattering, which is able to reduce the critical temperature in a two-gap superconductor.

The upper critical fields were analysed using the model proposed by Gurevich, which takes the multiband nature of superconductivity in $MgB_2$ into account. We observed how the scattering mechanisms determining critical field and resistivity values can be different. This explains why films with resitivities differing by one order of magnitude can show similar critical fields. $H_{c2}$ values up to 24 T in perpendicular direction and up to 57 T in the parallel orientation have been found. These high values confirm the importance of this material for large scale applications.

**Acknowledgments**

We acknowledge A.S. Siri, I. Pallecchi, A. Palenzona, R. Vaglio and L. Romanò for the helpul discussions. This work was supported by the European Community through "Access to Research Infrastructure action of the Improving Human Potential Programme".

|  | **FILM 1** | **FILM 2** | **FILM 3** | **FILM 4** |
|---|---|---|---|---|
| *substrate* | $Al_2O_3$ c-cut | MgO (111) | MgO (111) | $Al_2O_3$ c-cut |
| *c axis*, Å | 3.517 | 3.532 | 3.533 | 3.519 |
| $T_C$, K | 29.5 K | 32 K | 33.9 K | 38.8 K |
| $\Delta T_C$, K | 2.0 | 1.5 | 1.1 | 1.0 |
| RRR | 1.2 | 1.3 | 1.5 | 2.5 |
| $\rho(40K)$, $\mu\Omega$cm | 40 | 50 | 20 | 5 |
| $d\rho/dT(300\ K), (\mu\Omega\ cm/K)$ | 0.048 | 0.091 | 0.052 | 0.036 |

TABLE I. Main properties of the four thin films. The critical temperature value reported is the onset of the transition (90% of the normal state resistance) and the transition width is calculated between 90 % and 10 % of the normal state resistance. The absolute value of resistivity is with an accuracy of 20% due to the uncertainty in thickness determination. For comparison, the *c*-axis of the bulk is 3.521 Å.

|  | **FILM 1** | **FILM 2** | **FILM 3** | **FILM 4** |
|---|---|---|---|---|
| $\Gamma=\Gamma_{\sigma\sigma}\Gamma_{\pi\pi}/(\Gamma_{\sigma\sigma}+\Gamma_{\pi\pi})$, meV | 140 | 171 | 69 | 17 |
| $\Gamma_{\sigma\pi}$, meV | 4.2 | 3.2 | 2.4 | 0.1 |
| $D_{\sigma}$, $m^2 s^{-1}$ | $0.49\cdot 10^{-3}$ | $0.48\cdot 10^{-3}$ | $0.46\cdot 10^{-3}$ | $0.37\cdot 10^{-3}$ |
| $\tau$, s | $2.30\cdot 10^{-15}$ | $2.28\cdot 10^{-15}$ | $2.16\cdot 10^{-15}$ | $1.74\cdot 10^{-15}$ |
| $\rho_{\sigma}$, $\mu\Omega cm$ | 123 | 125 | 131 | 163 |
| $\rho_{\pi}$, $\mu\Omega cm$ | 59 | 83 | 23 | 5 |
| $\eta$ | 1.6 | 1.1 | 4.2 | 24 |
| $H_{c2}(0) \perp ab$, Tesla | 14.2 | 15.5 | 16.8 | 24.6 |
| $\gamma$ | 3.0 | 3.5 | 3.0 | 2.3 |
| $H_{c2}(0) //ab$, Tesla | 42 | 54 | 50 | 57 |

TABLE II. Some data drawn from resitivity curves and from critical field curves for the four films.

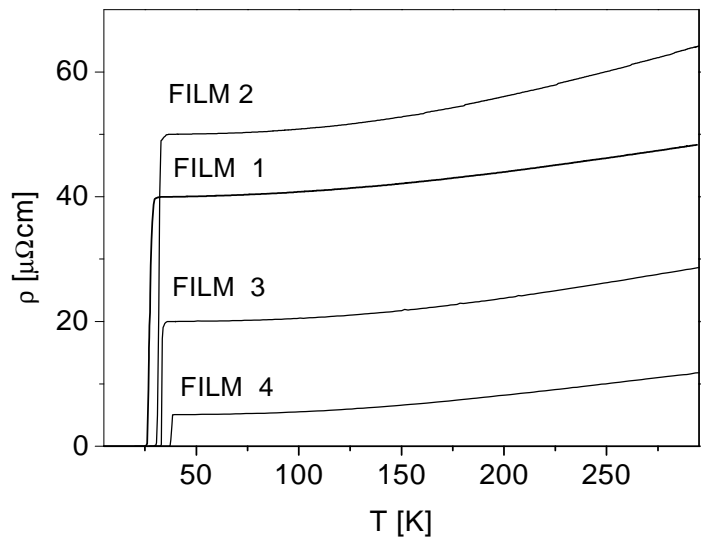

FIG.1. Resistivity as a function of temperature for the four films.

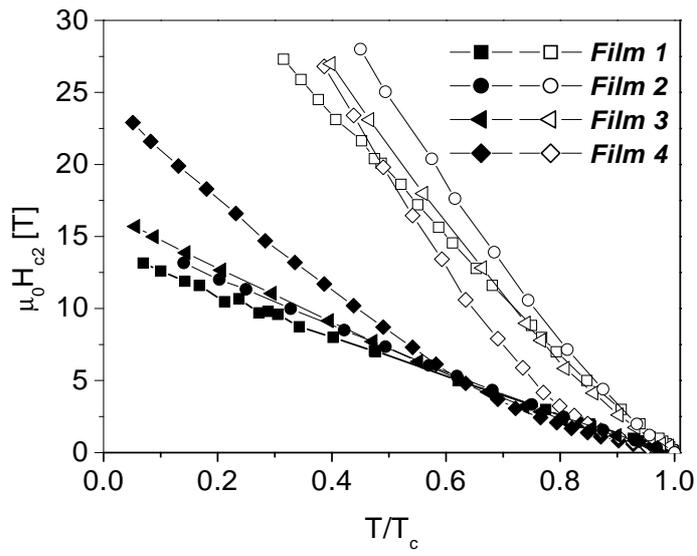

FIG. 2. Critical fields in the two orientations (parallel, open symbols, and perpendicular, full symbols, to the basal plane) for the four samples. For an easier comparison, they are presented as a function of the reduced temperature.

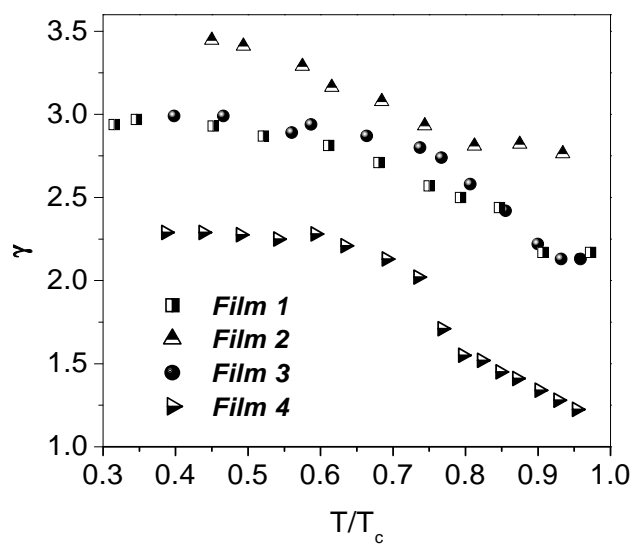

FIG. 3 Anisotropy factors $\gamma=H_{c2}(\theta=0°)/H_{c2}(\theta=90°)$ for all the films as a function of the reduced temperature.